\documentclass[final,5p,times,twocolumn]{elsarticle}

\usepackage{graphicx}

\usepackage{amssymb}

\journal{Physica C: Superconductivity and its application}

\begin{document}

\begin{frontmatter}


\title{Superconductivity of Cu/CuO$_{x}$ interface formed by shock-wave pressure}


\author[a]{A.~V.~Palnichenko\corref{cor1}}
\author[a]{N.~S.~Sidorov}
\author[b]{D.~V.~Shakhrai}
\author[b]{V.~V.~Avdonin}
\author[a]{O.~M.~Vyaselev}
\author[a]{S.~S.~Khasanov}
\address[a]{Institute of Solid State Physics, Russian Academy of Sciences,
Chernogolovka, Moscow region, 142432, Russia}
\address[b]{Institute of Problems of Chemical Physics, Russian Academy of Sciences,
Chernogolovka, Moscow region, 142432, Russia}

\cortext[cor1]{Corresponding author.  Tel.: +7 906 095 4402; fax: +7 496 524
9701. \textit{E-mail address}: paln@issp.ac.ru (A.~V.~Palnichenko)}

\begin{abstract}
A mixture of powdered Cu and CuO has been subjected to a shock-wave pressure of
$\simeq\,$350\,kbar with following quenching of the vacuum-encapsulated product
to $\approx 77\,$K. The ac magnetic susceptibility measurements of the samples
have revealed metastable superconductivity with $T_c \approx 19.5\,$K,
characterized by glassy dynamics of the shielding currents below $T_c$.
Comparison of the ac susceptibility and the dc magnetization measurements
infers that the superconductivity arises within the granular interfacial layer
formed between metallic Cu and its oxides due to the shock-wave treatment.

\end{abstract}

\begin{keyword}
Superconductivity\sep Copper-copper oxide interface\sep Dynamic magnetic
susceptibility\sep Superconductive glass


\end{keyword}

\end{frontmatter}


\section{Introduction}

Since the discovery of superconductivity above 30 K in lanthanum barium copper
oxide \cite{Bednorz}, many similar high-critical-temperature superconductors
(HTSs) have been found \cite{Plakida}. Superconductivity in all of these
multilayered oxides is believed to originate in the planes of copper and oxygen
atoms, common to these compounds \cite{Leggett}. Therefore, fabrication of
synthetic multilayers on the basis of the copper oxide layers is promising for
realization of new HTSs \cite{Triscone, Gariglio}. For example, a HTS-like
phenomenon was observed by means of electric conductivity measurements in the
samples consisting of Cu film deposited onto natural faces of CuO single
crystal under special conditions \cite{Osipov}. The phenomenon manifests itself
as a giant step-like rise of the electric conductivity of the samples by a
factor $\approx10^{5}$ as the temperature decreases below the critical value,
$T_c$, which varied in the range 200--500\,K depending on the thermal treatment
conditions of the samples. At $T<T_c$, the electric conductivity of the Cu/CuO
samples was found to be suppressed by applying the external magnetic field, as
well as by increasing of electric current in the sample. These experimental
facts suggest the superconductivity in the Cu/CuO interface with $T_c$
significantly higher than room temperature \cite{Osipov}.

A strong indication to superconductivity at 20--90\,K has been revealed by
dynamic magnetic susceptibility measurements in Cu/CuO$_x$ samples prepared by
surface oxidation of powdered copper followed by thermal treatment in vacuum
\cite{Sidorov_Cu}. By comparative dynamic and static magnetic susceptibility
measurements, the superconductivity was attributed to weakly coupled
superconducting islands formed spontaneously in the Cu/CuO$_x$ interfacial
areas during the heat treatment process.

Superconducting interfaces of such prepared Cu/CuO$_x$ samples are instable at
room temperature, as inferred from a decay of the superconducting transition
after a $\sim20$-hour exposure of the vacuum-encapsulated samples to room
temperature. However, quenching of the vacuum-encapsulated Cu/CuO$_x$ samples
to low temperature (77\,K) has helped to stabilize the superconducting
Cu/CuO$_x$ interfaces and prevent the decay of superconductivity
\cite{Sidorov_Cu}.

A success in stabilization of the superconducting Cu/CuO$_x$ interfaces by the
low temperature quenching has motivated our attempts to apply a shock-wave
pressure to the Cu/CuO mixture for their preparation. During the shock-wave
impact, the stroke energy applied to the sample evokes relative displacements
of local parts of the sample matter, resulting in a series of high-pressure
shock-waves propagating throughout the sample within $10^{-6} - 10^{-9}$\,s
\cite{Kanel}. The energy of the shock-waves leads to local, non-equilibrium
overheat of the sample's regions at the shock-wave front, followed by their
rapid cooling (quenching) as the shock-wave is passed, thus fixing the sample
in the metastable state. Furthermore, highly non-equilibrium conditions caused
by propagation of the high-pressure shock-waves in the sample can stimulate
phase transitions or mechanochemical reactions inaccessible by any equilibrium
processes, i.e. static pressure-temperature mode, resulting in new materials
\cite{Fortov, Sidorov_Ca}.

In this paper we report on metastable superconductivity at
$T_{c}\approx19.5$\,K revealed by the ac magnetic susceptibility measurements
of the powdered mixture of Cu and CuO subjected to shock-wave pressure of
$\approx350$\,kbar.

\section{Sample preparation and measurement techniques}

The samples were prepared by means of flat-type shock-wave pressure setup
described in detail earlier in Refs. \cite{Ossipyan, shock-wave-tchn}. The
starting samples were tablets, 9.6\,mm in diameter and 0.9\,mm thick, prepared
by pressing of 99.99\%-pure copper powder, 5--50\,$\mu$m grain size, covered by
0.2\,mm layer of powdered 10--40\,$\mu$m grain size, 99.99\%-pure copper oxide
CuO of monoclinic crystal structure (JCPDS 89-5899). For preparing the
superconducting Cu/CuO$_x$ samples, the optimum value of the shock-wave
pressure, within 1\,Mbar range, was found $\approx350$\,kbar.

After the shock-wave pressure treatment, the conservation cell was cut open and
the samples were extracted. Within 3--5\,min, the extracted samples were
vacuum-encapsulated to a residual pressure 1--5\,Pa into 5\,cm-long quartz
ampoules having 0.9\,mm-thick walls and 6\,mm outer diameter, and stored in
liquid nitrogen to prevent their degradation under normal conditions. We have
to note that quenching of the bare (vacuum-unsealed) shock-wave pressure
treated samples into liquid nitrogen did not result in superconductivity,
apparently due to destruction of the superconducting phase by adsorbed gaseous
layer on the sample surface during the sample cooling. In addition, for
comparative studies, the samples of compacted powdered pure Cu as well as pure
CuO and Cu$_{2}$O have been prepared separately under the same conditions as
described above.

The samples were studied by measuring the dynamic magnetic susceptibility,
 $\chi=\chi'-i\chi''$, using a mutual inductance ac susceptometer
\cite{shock-wave-tchn, Chen}. The amplitude $H_{ac}$ of the driving field
ranged from 0.22 to 12\,Oe, the driving frequency $\nu$ from 300\,Hz to
10\,kHz, and the superimposed dc magnetic field $H_{dc}$ up to 300\,Oe. In
order to prevent degradation, warming of the samples was avoided and the
measurements were done without unsealing the evacuated ampoule. The ampoule was
mounted on the measuring insert at $T\sim80$\,K, which was then dipped into a
precooled measurement cryostat.

Static magnetization of the sealed Cu/CuO$_x$ samples was studied by SQUID
magnetometer in the temperature range 5--70\,K and static magnetic fields 30--
300\,Oe. The ac susceptibility measurements have shown that the properties of
the sample sealed in the evacuated ampoule survive $\sim15$ minutes exposure to
room temperature, which enables standard loading routine of the SQUID
magnetometer.

Crystal structure of the samples was investigated in the temperature range
70--300\,K by X-ray diffraction measurements using Oxford Diffraction Gemini R
diffractometer, MoK$_\alpha$ radiation, equipped with a cooling system that
enables the measurements in the flow of cold nitrogen gas. For the diffraction
measurements, the sample was extracted from the quartz ampoule in the ambience
of liquid nitrogen and rapidly (within 5--10\,s) mounted onto the precooled
goniometer of the diffractometer.

\section{Experimental results}

\subsection{X-ray diffraction}

\begin{figure}[ht]
\center\includegraphics[scale=0.3,angle=90]{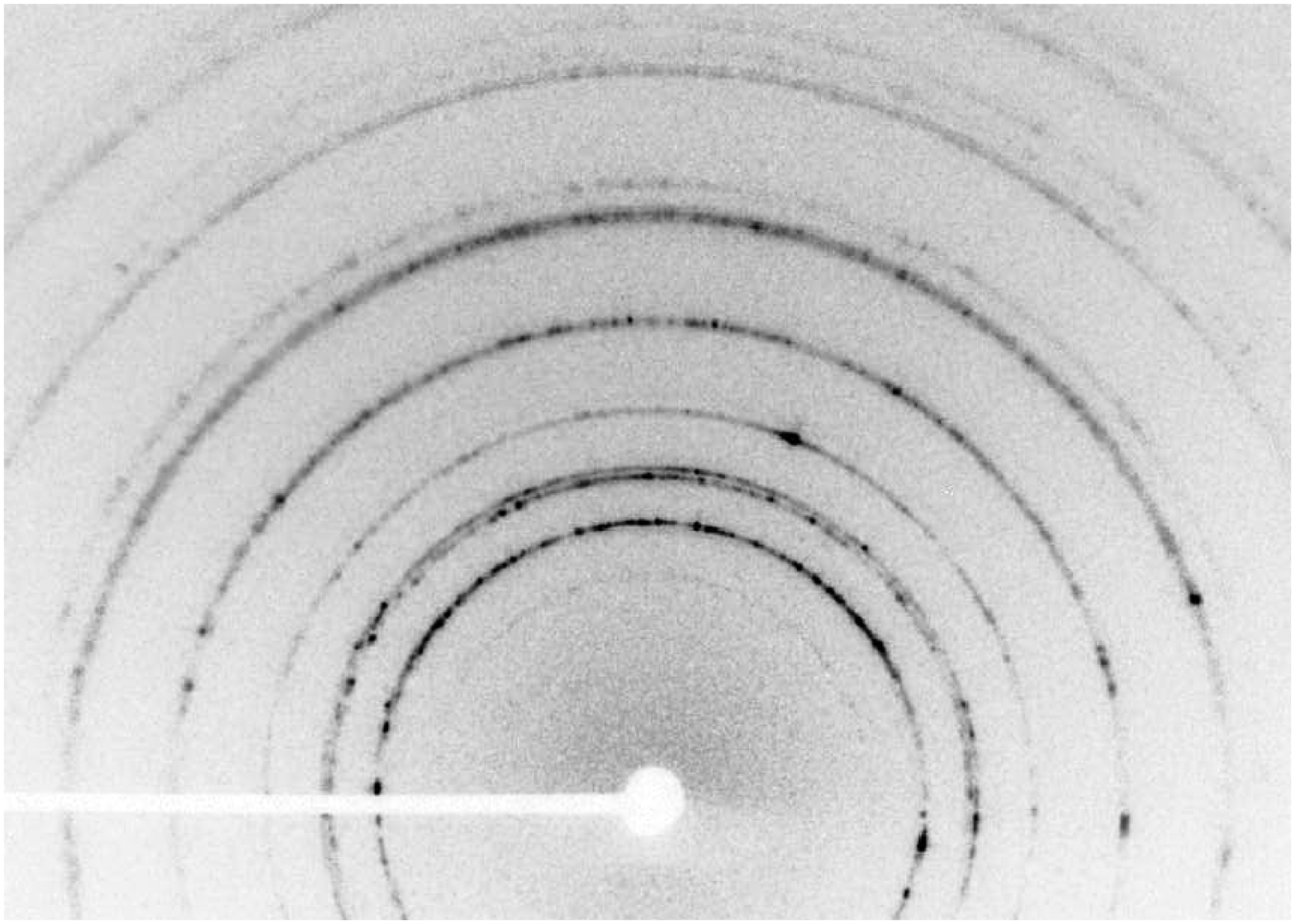} \caption{\label{Fig 1}
X-ray diffraction pattern for the shock-wave pressure treated Cu/CuO$_x$ sample
in MoK$_\alpha$ radiation. The complete diffraction ring set is a superposition
of the patterns from polycrystalline Cu, Cu$_2$O and CuO crystal structures.}
\end{figure}

Diffraction pattern of the shock-wave pressure treated Cu/CuO$_x$ sample,
recorded at $T\approx75$\,K, is shown in Fig.~\ref{Fig 1}. All the diffraction
rings in the observed pattern are a superposition of the patterns from
polycrystalline CuO ($a$ = 4.245\,\AA, space group Fm-3m), Cu$_2$O ($a$ =
4.267\,\AA, space group Pn-3m) and Cu ($a$ = 3.615\,\AA, space group Fm-3m)
crystal structures corrected by the thermal expansion factor \cite{ICDD}. No
change has been detected in the X-ray diffraction patterns of the Cu/CuO$_x$
samples after their exposure for $\approx20$ hours at room temperature.

\subsection{Dynamic susceptibility}

\begin{figure}[ht]
\center\includegraphics[scale=0.3]{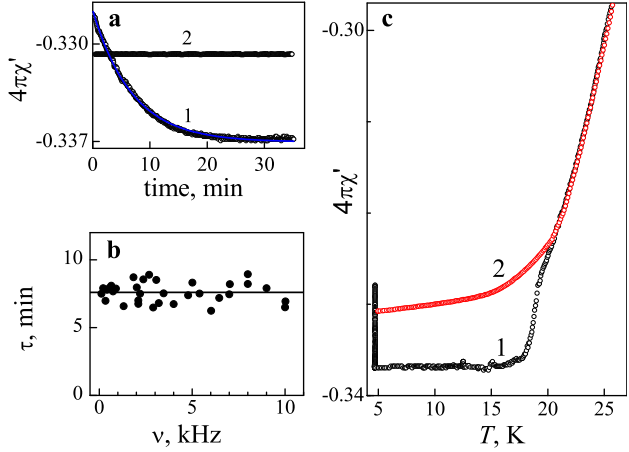} \caption{\label{Fig 2}(a) Time
evolution of $4\pi\chi'$ at $T$ = 4.7\,K for the vacuum-encapsulated shock-wave
treated Cu/CuO$_x$ sample (Curve~1) and for the same sample exposed to room
temperature for $\sim20$ hours (Curve~2). Solid curve: fit to Curve~1 in the
form $a + b\exp(-t/\tau)$, where \textit{a}, \textit{b} and $\tau$ are fit
parameters. The measurements were performed at $\nu$ = 2.1\,kHz, $H_{ac}$ =
0.6\,Oe, $H_{dc}$ = 0.(b) The dependence of the time constant $\tau$ on drive
frequency $\nu$ ($H_{ac}$ = 0.6\,Oe, $H_{dc}$ = 0). (c) Temperature
dependencies of $4\pi\chi'(T)$ measured before (Curve~1) and after (Curve~2)
the sample exposure to room temperature ($\nu$ = 2.1\,kHz, $H_{ac}$ = 0.6\,Oe,
$H_{dc}$ = 0).}
\end{figure}

Each cycle of $\chi(T)$ measurements started from a cooldown of the Cu/CuO$_x$
sample to 4.7\,K. Next, $H_{ac}$ was switched onto enable the measurements.
Quite remarkably, at constant $T$ = 4.7\,K, $\chi'$ develops to the equilibrium
value monotonously in time, $t$, towards the enlargement of diamagnetism.
Curve~1 in Fig.~\ref{Fig 2}a illustrates the time evolution of $\chi'$ measured
at frequency  $\nu$ = 2.1\,kHz, $H_{ac}$ = 0.6\,Oe and $H_{dc}$ = 0. $\chi'(t)$
was found to relax exponentially to ground diamagnetic state as $a +
b\exp(-t/\tau)$, shown in Fig.~\ref{Fig 2}a by a solid line, where \textit{a},
\textit{b} and $\tau$ are fit parameters. The extracted time constant,
$\tau=7.6\pm1.4\,$min, is practically flat in frequency for the range 0.1 -
10\,kHz, as shown in Fig.~\ref{Fig 2}b. We have to note that the observed
$\chi'(t)$ dependencies were not influenced by an effect of the sample
temperature relaxation towards the equilibrium temperature, because the
$\chi'(t)$ relaxation curves were found to be insensitive to the retention
interval of the sample exposure at $T = 4.7$\,K ($H_{ac} = 0$) until the
measurements start.

After a half-hour ($\sim4\tau$) delay at $T$ = 4.7\,K ($H_{ac}$ = 0.6\,Oe,
$H_{dc}$ = 0), $\chi(T)$ dependence was measured upon heating at the rate of
1--1.5\,K/min (slower heating rate made no visible change to the measurement
result). The measured $\chi'(T)$ dependence is shown by Curve~1 in
Fig.~\ref{Fig 2}c. In this plot, the drop in $\chi'(T)$ curve at 4.7\,K
corresponds to the $\chi'(t)$ relaxation process, Curve~1 in Fig.~\ref{Fig 2}a.
As the temperature increases, a step-like rise of $\chi'(T)$ is observed at
$T_c\approx18.5$\,K ($H_{ac}$ = 0.6\,Oe, $H_{dc}$ = 0\,Oe), signifying a phase
transition in the sample at this temperature. In order to exclude the influence
of the $\chi'(t)$ relaxation process on the $\chi(T)$ result, all subsequent
$\chi(T)$ measurements were performed according to the measurement cycle
described above (cooldown to 4.7\,K - turn on $H_{ac}$, $H_{dc}$ - half-hour
delay - sweep up $T$ and measure).

\begin{figure}[ht]
\center\includegraphics[scale=0.4]{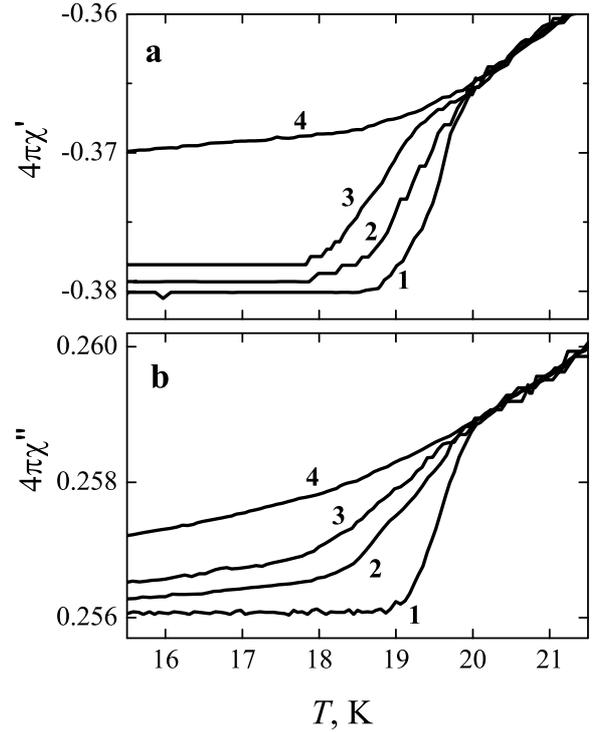} \caption{\label{Fig 3}
Temperature dependencies of the real (a) and the imaginary (b) parts of
$4\pi\chi$  for the Cu/CuO$_x$ sample measured under the dc magnetic field
$H_{dc}$ = 0, 168\,Oe and 268\,Oe (Curves~1 to 3, respectively). Curves~4
correspond to the same sample exposed for $\approx20$ hours to room
temperature. The measurements were performed at $\nu$ = 2.94\,kHz in $H_{ac}$ =
0.22\,Oe. }
\end{figure}

\begin{figure}[ht]
\center\includegraphics[scale=0.4]{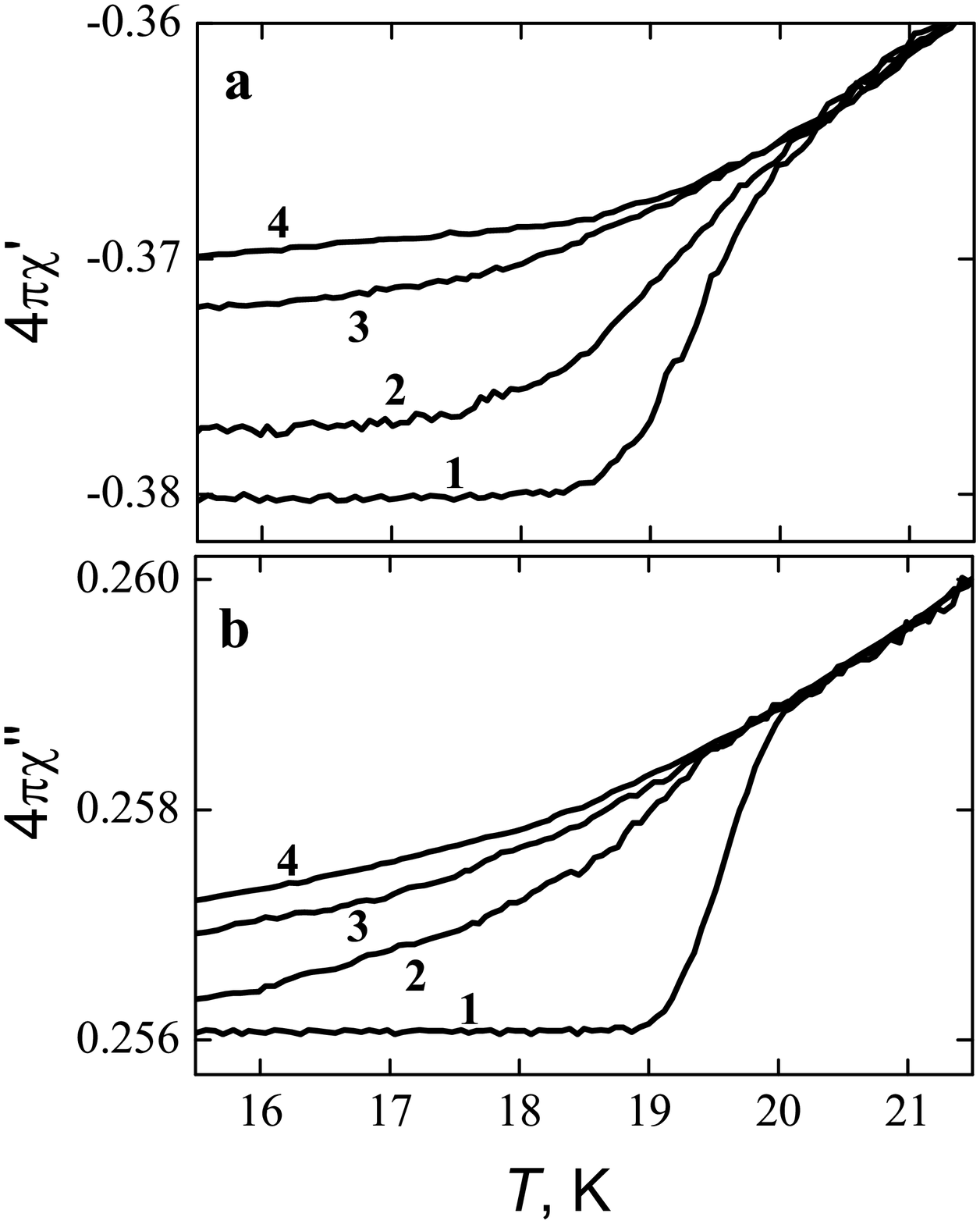} \caption{\label{Fig 4}
Temperature dependencies of the real (a) and the imaginary (b) parts of
$4\pi\chi$  for the Cu/CuO$_x$ sample measured with the amplitude of the ac
driving field $H_{ac}$ = 0.22, 1.8 and 11.6\,Oe (Curves~1 to 3, respectively).
Curves~4 correspond to the same sample exposed for $\approx20$ hours to room
temperature. The measurements were performed at $\nu$ = 2.94\,kHz in $H_{dc}$ =
0.}
\end{figure}

The $\chi(T)$ dependencies measured in a static magnetic field, $H_{dc}$, are
shown in Fig.~\ref{Fig 3}. According to Curves~1-3, an increase in $H_{dc}$
suppresses the anomaly in $\chi(T)$. Increasing the driving ac magnetic field
amplitude, $H_{ac}$, gives a similar effect, as shown in Fig.~\ref{Fig 4}.

Fig.~\ref{Fig 5} shows the temperature dependencies of $\chi'$ and $\chi''$
measured at frequencies from 0.3 to 10\,kHz. One can see in Fig.~\ref{Fig 5}
that $T_c$ is essentially frequency independent, while the shapes of the
$\chi'(T)$ and $\chi''(T)$ curves at the phase transition change dramatically
with the frequency.

After completion of these measurements, the insert with the Cu/CuO$_x$ sample
was finally kept for $\approx20$ hours at room temperature. The following
measurements have shown, first of all, no time dependence of $\chi'(t)$ at
4.7\,K, see Curve~2 in Fig.~\ref{Fig 2}a. Furthermore, the $\chi(T)$ dependence
of the room-temperature-annealed sample does not exhibit a step-like anomaly
and is independent on $H_{ac}$ and $H_{dc}$, within the ranges used previously.
Moreover, now $\chi'(T)$ and $\chi''(T)$ dependencies (Curve~2 in Fig.~\ref{Fig
2}c, Curves~4 in Figs.~\ref{Fig 3},~\ref{Fig 4}) are practically identical to
those of pure powdered copper sample, subjected to the shock-wave pressure
treatment under the same conditions. Next, $\chi(T)$ of the shock-wave pressure
treated pure copper oxide (CuO, Cu$_2$O) samples did not reveal any anomaly in
the experimental temperature range. Thus all the measuring results denote that
the phase responsible for the observed anomaly in $\chi(T)$ of the Cu/CuO$_x$
samples is related to the interfacial area, formed between the copper and
copper oxide polycrystalline phases, which is unstable at room temperature.

\subsection{dc magnetic moment}

In contrast to the ac magnetic susceptibility, the dc magnetic moment of the
shock-wave pressure treated Cu/CuO$_x$ samples, measured using a SQUID
magnetometer  in a standard zero-field-cooled and field-cooling regimes,
revealed no sign of the magnetic anomaly. At temperatures from 5 to 70\,K, the
samples demonstrated a nearly temperature-independent magnetic susceptibility
($\sim10^{-6}$\,cm$^{3}$/g) in fields $H_{dc}$ = 30--300\,Oe.

\section{Discussion}

\subsection{Evidence of metastable superconductivity}

In the ac magnetic susceptibility measurements, the $\chi'$ response to the ac
magnetic field is a function of the skin depth,
$\delta\propto1/\sqrt{\sigma\mu\nu}$, where $\sigma$ and $\mu$ are,
respectively, the electric conductivity and magnetic permeability of the
material, and  $\nu$ is the ac magnetic field frequency. According to the dc
magnetic moment measurements, the Cu/CuO$_x$ samples are nonmagnetic
($\mu\approx 1$). Moreover, no anomalies in the dc magnetic moment have been
found in the range 5--70\,K. Therefore, the only reason for the observed
step-like anomaly in $\chi'(T)$, shown by Curve~1 in Fig.~\ref{Fig 2}c, is
sudden change of the electric conductivity of the sample, which we attribute to
a superconducting transition.

In order to verify this assumption, the influence of magnetic field on the
anomaly in $\chi(T)$ has been studied (Figs.~\ref{Fig 3},~\ref{Fig 4}). The
anomaly in $\chi(T)$ observed at $T_c\approx19.5$\,K for $H_{ac}$ = 0.22\,Oe
and $H_{dc}$ = 0, reduces in size and shifts to lower temperature with
increasing either $H_{dc}$ (Curves~1--3 in Fig.~\ref{Fig 3}) or $H_{ac}$
(Curves~1--3 in Fig.~\ref{Fig 4}). The $\chi'(T)$ and $\chi''(T)$ dependencies
presented in Figs.~\ref{Fig 3},~\ref{Fig 4} are typical for superconducting
materials \cite{Hein, Ishida} which implies a superconducting transition in the
Cu/CuO$_x$ sample.

\begin{figure}[ht]
\center\includegraphics[scale=0.45]{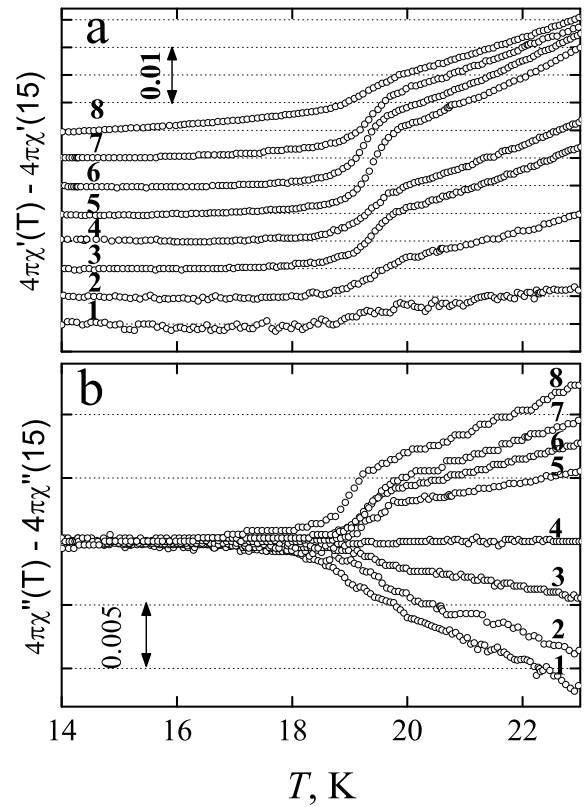} \caption{\label{Fig 5}
Temperature dependencies of the real (a) and imaginary (b) parts of $4\pi\chi$
for the Cu/CuO$_x$ sample measured at frequencies 0.313, 0.93, 1.53, 2.1, 2.9,
3.46, 5.4 and 10.26\,kHz (Curves~1 to 8, respectively). $H_{ac}$ = 0.22\,Oe,
$H_{dc}$ = 0. For visual clearness, the $4\pi\chi'(T)$ curves are separated
along Y-axis by 0.005.}
\end{figure}

It would be worthful to note that none of the constituents of the sample,
neither metallic Cu nor Cu$_x$O (x = 1, 2) are superconductors in the bulk.
Moreover, none of them, taken separately and subjected to the shock-wave
pressure treatment under the same conditions, demonstrate any anomaly in
$\chi(T)$. We conclude therefore that the superconductivity in the shock-wave
treated Cu/CuO$_x$ sample is related to the interfacial layer formed between
copper oxide and metallic Cu phases.

One can see in Figs.~\ref{Fig 3},~\ref{Fig 4} that at temperatures well below
$T_c$, $4\pi\chi'(T)>-1$ and $\chi''(T)> 0$, whereas a bulk superconductor
assumes an ideal diamagnetic state with $4\pi\chi'(T)=-1$ and $\chi''(T)= 0$.
Besides, the zero-field-cooled regime of the dc magnetization measurement has
shown no sign of the superconducting transition at all. We conclude therefore
that the superconducting phase formed within the interfacial layer is not a
closed surface enveloping the sample, which would trap the magnetic flux and
keep it constant, but rather consists of superconducting grains embedded into a
non-superconducting host matrix. The grains are apparently small, less than the
London penetration depth, $\lambda$, and are weakly linked by narrow,
non-superconducting bridges.

In this assumption, in the dc magnetization measurements, the weak
non-superconducting links result in rapid decay of the intergrain shielding
current after applying the dc magnetic field, while the intragrain
superconducting currents give a negligible contribution to the shielding of
magnetic field due to small (sub-$\lambda$) size of the grains, thus resulting
in nonmagnetic ($\mu\sim1$) superconducting state of the sample.

On the contrary to the dc magnetization measurements, the alternating magnetic
field used in the ac magnetic susceptibility measurements maintains the
intergrain shielding current. This enables the superconductivity to be detected
even in discontinuous, energy dissipative loops, consisting of the weakly
linked superconducting sub-$\lambda$-grains, although the diamagnetic state of
such structures is quite obviously far from ideal, as observed for the
Cu/CuO$_x$ system.

As one can see in Fig.~\ref{Fig 5}a, the step in $\chi'(T)$ at
$T_c\approx19.5$\,K becomes bigger as the driving frequency increases from
0.313 to 2.9\,kHz (Curves~1--5) and diminishes with further frequency increase
(Curves~6--8). This can be explained by a superposition of normal and
superconducting shielding currents, contributing to $\chi'(T)$ in the
Cu/CuO$_x$ sample.

Consider first the $\chi'(T)$ response at $\approx19.5$\,K, diminishing with
the frequency increase ($\nu\geq2.9$\,kHz, Curves~5--8 in Fig.~\ref{Fig 5}a).
This effect is well known for bulk metallic superconductors: in the normal
state, increasing the ac magnetic field frequency leads to diminishing skin
depth $\delta$, hence a bigger diamagnetic signal. As a result, in the high
frequency limit, the normal metal just above $T_c$ is nearly as diamagnetic as
the superconductor just below $T_c$ which makes the superconducting transition
indistinguishable in $\chi(T)$.

On the other hand, intergranular weak links in the superconducting Cu/CuO$_x$
interfaces can be considered as electric capacitors connected in series into
the shielding intergrain supercurrent loops. Obviously, the higher is
frequency, the better the superconducting grains are linked, hence the stronger
is the superconducting shielding. Apparently, this explains the growth of the
$\chi'(T)$ response at $T_c\approx19.5$\,K at frequencies up to 2.9\,kHz
(Curves~1--5 in Fig.~\ref{Fig 5}a). The two competing contributions superimpose
in the studied Cu/CuO$_x$ sample to give a maximum of the $\chi'(T)$ response
to the superconducting transition at $\nu = 2.9$\,kHz.

The evolution of $\chi''(T)$ with frequency is illustrated in Fig.~\ref{Fig
5}b. At high frequencies (Curves~5--8), $\chi''(T)$ exhibits a step-like drop
at cooling below $T_c\approx19.5$\,K, which reflects shielding of the
energy-dissipative interior of the Cu/CuO$_x$ sample by the intergrain
supercurrents. The drop is vanished at low frequencies (Curves~1--4),
apparently due to decreasing of the superconducting intergrain shielding at low
frequencies.

As it was mentioned above, the superconducting interfaces in the Cu/CuO$_x$
sample are unstable at room temperature (Figs.~\ref{Fig 2},~\ref{Fig 3} and
\ref{Fig 4}). The reason for such instability is apparently the copper/oxygen
ionic diffusion processes activated at room temperature in the Cu/CuO$_x$
interface, which destroy the superconductivity in the sample.

The $T$-dependencies of $\chi'$ and $\chi''$ measured on the degraded
Cu/CuO$_x$ samples (Curves~4 in Figs.~\ref{Fig 3},~\ref{Fig 4}) are typical for
normal nonmagnetic metals. Namely, as the temperature decreases, the electric
conductivity of the sample increases monotonically that leads to a decrease in
the skin depth $\delta$ , thus to a monotonic decrease in $\chi'$
\cite{Landau}. At low temperatures, the electric conductivity becomes more flat
in temperature, and $\chi'(T)$ sets constant. In turn, $\chi''$ is also a
function of $\delta$: monotonous increase of $\chi''$ with temperature
corresponds to an increase of $\delta$ while $\delta < r$, where $r$ is the
sample depth. As the temperature continues to increase, $\chi''(T)$ reaches a
maximum at $T_{max}$, corresponding to $\delta\approx r$, followed by
$\chi''(T)$ decreasing at $\delta > r$ \cite{Landau}. Therefore, an increase of
the ac magnetic field frequency $\nu$ moves the $T_{max}$ to the higher
temperatures, resulting in change of the $\chi''(T)$ slope with the frequency
at $T>T_c$, observed in Fig.~\ref{Fig 5}b.

\subsection{Superconductive glass}

In granular superconductors, the superconducting grains are weakly coupled into
closed loops supporting the shielding supercurrents in response to an external
ac magnetic field. Due to the positional disorder of individual superconducting
grains, the supercurrent loops can form multiple states of nearly equal energy.
However, energy barriers between these states tend to inhibit hops from one
energy state to another, resulting in time-dynamics of the frustrated shielding
supercurrents similar to that observed in spin glasses \cite{Young, Ebner,
Blatter}.

In our experiments, the glassy behavior of the supercurrent loops in the
Cu/CuO$_x$ sample is evident from the observed slow $\chi'(t)$ relaxation into
the lower energy diamagnetic ground state at 4.7\,K (Fig.~\ref{Fig 2}a). The
relaxation time, $\tau$ = 7.6\,min, was found frequency independent in the
range 300\,Hz--10\,kHz. The insensibility of $\tau$ to the excitation frequency
$\nu$ is presumably a consequence of a significant time-scale mismatch between
the $\chi'(t)$ relaxation process and the ac magnetic field oscillation period,
$\tau\gg1/2\pi\nu$. This result is qualitatively similar to that for the
shock-wave pressure treated Mg/MgO samples \cite{shock-wave-tchn}.  The origin
of such a long $\chi'(t)$ relaxation process is rather puzzling. No clear
explanation for the observed dynamics of $\chi'(t)$ in a weak ac magnetic field
has been found on the basis of existing approaches that involve collective
creep or vortex-glass theories for granular superconductors \cite{Blatter}.

\subsection{Nature of the Observed Superconductivity}

Although the room-temperature exposure of the Cu/CuO$_x$ samples leads to a
dramatic decay of the superconducting fraction, no corresponding change has
been detected in the X-ray diffraction patterns (Fig.~\ref{Fig 1}). The
fraction of the superconducting phase is therefore too small to be detected by
the X-ray diffraction technique. In addition, none of the constituents of the
sample, neither metallic Cu nor Cu$_x$O (x = 1, 2), taken separately and
subjected to the shock-wave pressure treatment under the same conditions,
demonstrate any anomaly in $\chi(T)$. We believe therefore that the
superconductivity in the shock-wave pressure treated Cu/CuO$_x$ samples is
related to the interfacial layer formed between copper oxide and metallic
copper phases.

One may reasonably ascribe the superconductivity in the Cu/CuO$_x$ samples to
some unknown superconducting copper oxide structure, unstable under the normal
conditions and located at the interfacial areas in the Cu/CuO$_x$ sample. For
example, we could suppose a two-dimensional CuO lattice which is formed on the
surface of CuO in the Cu/CuO  interfaces. This lattice, consisting of Cu$^{2+}$
and O$^{1-}$ ions, may form a narrow, partially filled two-dimensional band. In
this case, the onset temperature of Bose-Einstein condensation may take a value
of $T_c\sim1000$\,K \cite{Amelin}.

However, the superconductivity has been discovered for other similar objects
based on metals of various groups and their oxides (Mg/MgO
\cite{shock-wave-tchn, Sidorov_Mg}, Na/NaO$_x$ \cite{Sidorov_Na},
Al/Al$_2$O$_3$ \cite{Sidorov_Al} and Fe/FeO$_x$ \cite{Sidorov_Fe}). This
indicates the generality of the observed phenomenon, related to the oxygen
state at the interface, rather than a unique property of Cu/CuO$_x$ system
\cite{Amelin}.

Alternatively, consider isolated nanometer-sized metallic clusters which may
spontaneously arise in the granular metal-oxide interfacial areas, formed
during the shock-wave pressure \cite{shock-wave-tchn} or the surface oxidation
processes \cite{Sidorov_Cu, Sidorov_Mg, Sidorov_Na, Sidorov_Al, Sidorov_Fe}.
Delocalized electrons of such cluster are expected to form narrow, partially
filled energy band at the Fermi level, resulting in a new hypothetical family
of high-T$_c$ ($>100$\,K) superconductors \cite{Friedel, Kresin, Ovchinnikov,
Cao}.

Besides, two-dimensional metal/oxide interfacial areas, formed in the
Cu/CuO$_x$ sample, may play a role of asymmetric confining potentials in the
system of free electrons, resulting in the lack of spatial inversion symmetry
at the interfaces, which may stimulate to a topological change of the Fermi
surface, due to the spin-orbit splitting, and lead to the enhanced
superconductivity \cite{Cappelluti}.

\section*{Acknowledgement}

We gratefully acknowledge useful discussions of the results with
V.~V.~Ryasanov. The work has been supported by RAS Presidium Programs "Quantum
physics of condensed matter" and "Thermal physics and mechanics of extreme
energy impacts and physics of strongly compressed matter" as well as RFBR grant
No.~13-02-01217a .

\section*{References}

\bibliographystyle{model1-num-names}
\bibliography{Cu}

\end{document}